\documentclass[conference]{IEEEtran}
\usepackage{cite}

\ifCLASSINFOpdf
  \usepackage[pdftex]{graphicx}
\usepackage{tikz}
\usepackage{graphicx}
\usepackage{epstopdf}
\usepackage{pgf}

  \graphicspath{{../}}
\else
\fi
\usepackage[cmex10]{amsmath}
\usepackage{amssymb}
\usepackage{amsthm}

\usepackage{algorithmic}

\usepackage{array}

\usepackage{mdwmath}
\usepackage{mdwtab}

\usepackage[tight,footnotesize]{subfigure}
\begin{document}

\title{Optimal Signaling of MISO Full-Duplex Two-Way Wireless Channel}
\author{\IEEEauthorblockN{Shuqiao Jia and Behnaam Aazhang}
\IEEEauthorblockA{Department of Electrical and Computer Engineering\\
Rice University\\
Houston, Texas 77005\\
Email: \{shuqiao.jia, aaz\} @rice.edu}
}


%


\maketitle

\begin{abstract}
We model the self-interference in a multiple input single output (MISO) full-duplex two-way channel and evaluate the achievable rate region. We formulate the boundary of the achievable rate region termed as the Pareto boundary by a family of coupled, non-convex optimization problems. Our main contribution is decoupling and reformulating the original non-convex optimization problems to a family of convex semidefinite programming problems. 
For a MISO full-duplex two-way channel, we prove that beamforming is an optimal transmission strategy which can achieve any point on the Pareto boundary. Furthermore, we present a closed-form expression for the optimal beamforming weights. In our numerical examples we quantify gains in the achievable rates of the proposed beamforming over the zero-forcing beamforming.
\end{abstract}


%

\section{Introduction}

A node in full-duplex mode can simultaneously transmit and receive in the same frequency band. Therefore, the wireless channel between two full-duplex nodes can be bidirectional, having the potential to double the spectral efficiency when compared to the half-duplex network. Due to the proximity of the transmitters and receivers on a node, the overwhelming self-interference becomes the fundamental challenge in implementing a full-duplex network. The mitigation of the self-interference signal can be managed at each step of the communication network by passive and active cancellation methods \cite{sahai2013ITVT}. In recent results \cite{Everett2011Allerton,Everett2011Asilomar,bharadia2013sigcomm}, the feasibility of the single input single output (SISO) full-duplex communication has been experimentally demonstrated. However, the performance is limited by the residual self-interference which is considered in \cite{sahai2013ITVT,bharadia2013sigcomm,vehkapera2013PIMRC,Day2012ITSP} to be induced by the imperfection of the transmit front-end chain.

The bottleneck from imperfect transmit front-end chain has motivated recent research in full-duplex channel with transmit front-end noise. The performance of the SISO full-duplex two-way channel has been thoroughly analyzed in \cite{Duarte2012ITWC,sahai2013ITVT}. The multiple input multiple output (MIMO) full-duplex two-way channel with transmit front-end noise is considered in \cite{vehkapera2013PIMRC,Day2012ITSP} (in \cite{vehkapera2013PIMRC,Day2012ITSP} termed as MIMO full-duplex bidirectional channel). In \cite{vehkapera2013PIMRC} Vehkapera \textit{et al.} studied the effect of time-domain cancellation and spatial-domain suppression on the channel, while in \cite{Day2012ITSP} Day \textit{et al.} derived the lower bound of achievable sum-rate for the channel and proposed a numerical search for optimal signaling. In this paper, we focus on the multiple input single output (MISO) full-duplex two-way channel in presence of the transmit front-end noise. Compared with \cite{Day2012ITSP}, we derive the tight boundary of the achievable rate region for the channel and present the analytical closed-form solution of the optimal signaling. Note that the achievable rate region includes the achievable sum-rate as a point and provides the additional asymmetric performance metric. 

\begin{figure}[!t] 
\centering
\includegraphics[width=2.8in]{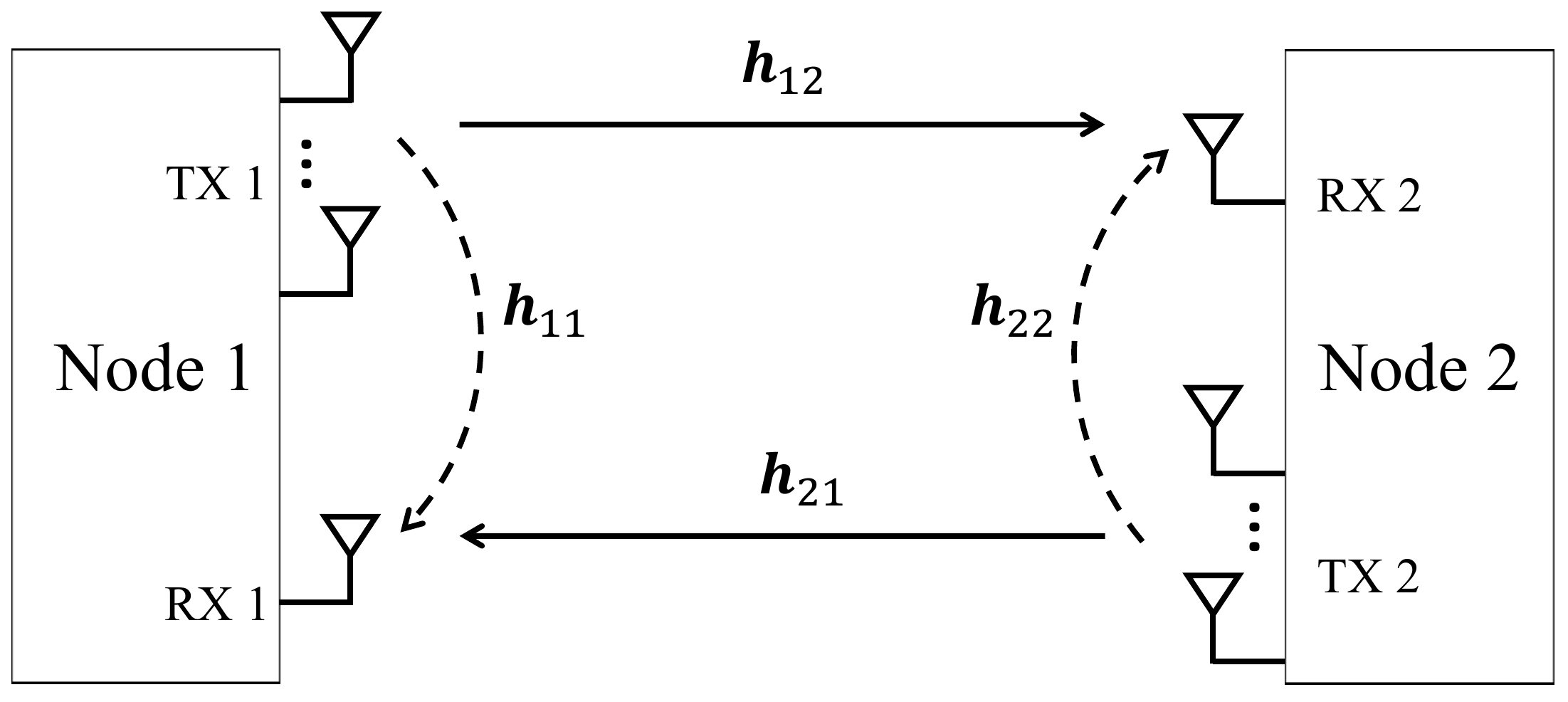}
\caption{The MISO point-to-point full-duplex network under study. The solid line denotes the desired channel and the dashed line denotes the self-interference channel.} 
\label{Fig1}
\end{figure}

In this paper, we consider the optimal signaling structure for the MISO full-duplex two-way channel, by which all rate pairs on the boundary of the achievable rate region can be achieved. We introduce the channel model that includes transmit front-end noise. We leverage our model to characterize the achievable rate region for the full-duplex channel. The boundary of the region is described by a family of non-convex optimization problems. Rendering the computation tractable, we decouple the original non-convex problems to the family of convex optimization problems. The decoupling method was first developed in the field of game theory \cite{osborne2004gametheory} and recently introduced to communications in \cite{Shang2007Asilomar,Shang2011ITInf,Larsson2008JSAC,jorswieck2008ITSP,zhang2010ITSP}. By employing the semi-definite programing (SDP) reformulation, we numerically solve the optimal signaling and prove the optimality of transmit beamforming. That is to say, for a MISO full-duplex two-way channel, all the points on the boundary of the achievable rate region can be achieved by restricting to transmit beamforming scheme. Furthermore, we derive the closed-form optimal beamforming weights. Finally, through simulations we show the achievable rate regions for the MISO full-duplex two-way channels and evaluate the performance of the traditional zero-forcing beamforming with our optimal beamforming.

\textit{Notation}: We use $(\cdot)^{\dag}$ to denote conjugate transpose. For a scalar $a$, we use $|a|$ to denote the absolute value of $a$. For a vector $\boldsymbol{a}\in \mathbb{C}^{M\times 1}$, we use $\|\boldsymbol{a}\|$ to denote the norm, $\boldsymbol{a}^{(k)}$ to denote the $k^{th}$ element of $\boldsymbol{a}$, $\text{Diag}(\boldsymbol{a})$ to denote the square diagonal matrix with the elements of vector $\boldsymbol{a}$ on the main diagonal. For a matrix $\boldsymbol{A}\in \mathbb{C}^{M\times M}$, we use $\boldsymbol{A}^{-1}$, $\textbf{tr}(\boldsymbol{A})$ and $\text{rank}(\boldsymbol{A})$ to denote  the inverse, the trace and the rank of $\boldsymbol{A}$, respectively. We use $\text{diag}(\boldsymbol{A})$ to denote the diagonal matrix with the same diagonal elements as $\boldsymbol{A}$. $\boldsymbol{A}\succeq 0$ means that $\boldsymbol{A}$ is a positive semidefinite Hermitian matrix. We denote expectation, variance and covariance by $\text{E}\{\cdot\}$, $\text{Var}\{\cdot\}$ and $\text{Cov}\{\cdot\}$, respectively. Finally, $\mathbb{C}$ and $\mathbb{H}$ denotes the complex field and the Hermitian symmetric space, respectively.

\section{Channel Model}
We present the channel model for a MISO full-duplex network with two nodes as illustrated in Fig. \ref{Fig1}. Assume two nodes indexed by $i,j\in\{1,2\}$ share the same single frequency band for transmission. Each node is equipped with $M$ transmit antennas and a single receive antenna. The signal from transmitter $i$ is collected as the signal of interest by receiver $j, j\neq i$, while appears at its own receiver $i$ as the self-interference signal.

\begin{figure}[!t] 
\centering
\includegraphics[width=3.1in]{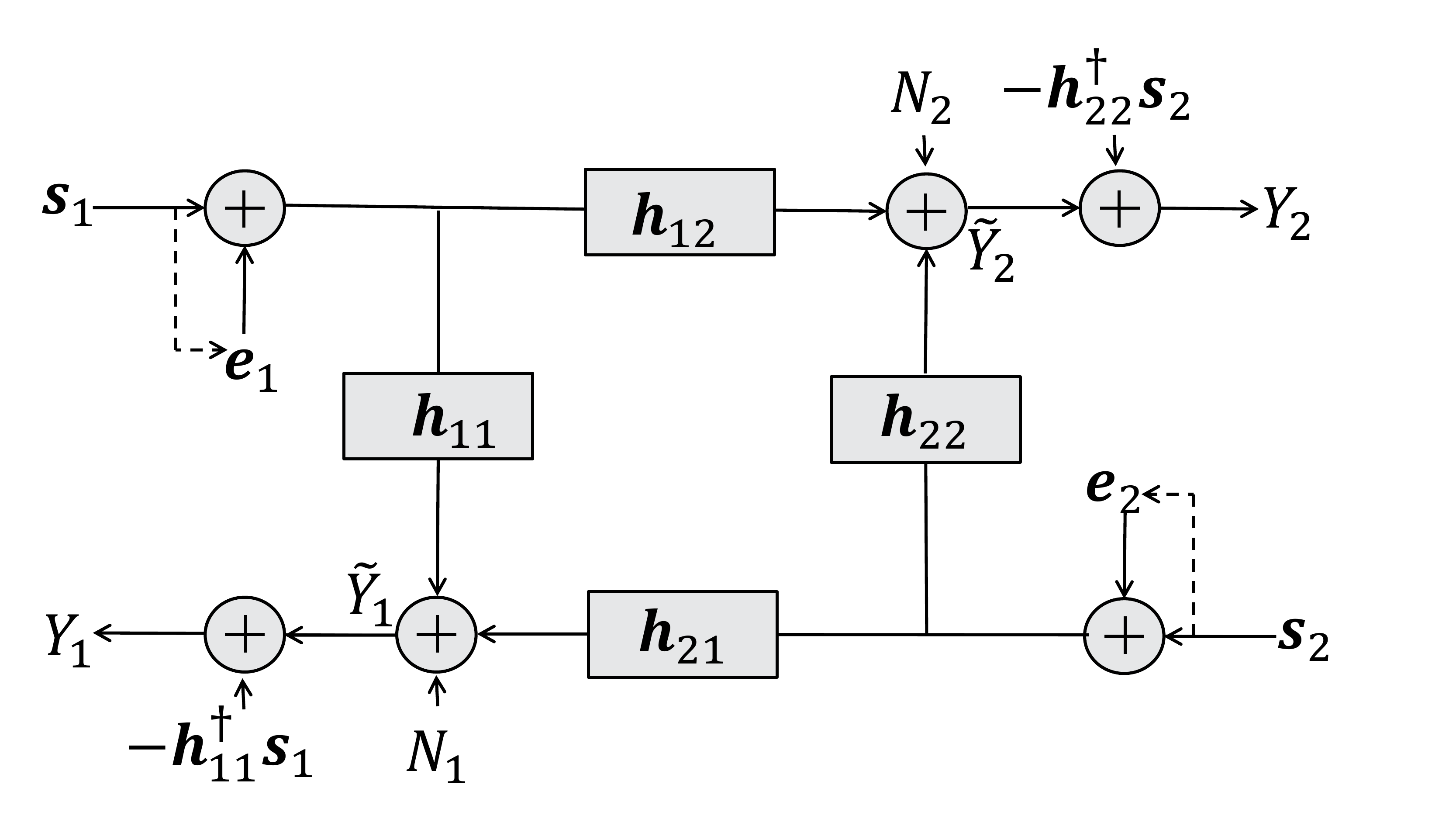}
\caption{The model of the MISO full-duplex two-way channel under study.} 
\label{Fig2}
\end{figure} 
             
Fig. \ref{Fig2} summarizes our model of the MISO full-duplex two-way channel. Denote the wireless channel from transmitter $i$ to receiver $j$ by the complex vector $\boldsymbol{h}_{ij}\in \mathbb{C}^{M\times 1}$. The signal at receiver $i$ is given by
\begin{equation}
    \widetilde{Y}_i=\boldsymbol{h}_{ji}^{\dag}(\boldsymbol{s}_j+\boldsymbol{e}_j)+\boldsymbol{h}_{ii}^{\dag}(\boldsymbol{s}_i+\boldsymbol{e}_i)+N_i
   \label{received signal}
   \end{equation}  
where $\boldsymbol{s}_i\in \mathbb{C}^{M\times1}$ denotes the transmit signal prior to the transmit front-end chain at transmitter $i$. An additional transmit front-end noise $\boldsymbol{e}_i$ is propagated over the same channel as $\boldsymbol{s}_i$. At receiver $i$, the thermal noise is modeled as a complex Gaussian noise $N_i\sim\mathcal{CN}(0,\sigma_N^2)$.

The transmit front-end noise $\boldsymbol{e}_i$ is induced by the imperfect transmit front-end chain \cite{bharadia2013sigcomm,vehkapera2013PIMRC}. More precisely $\boldsymbol{e}_i$ statistically relates to the transmit signal $\boldsymbol{s}_i$ due to the limited dynamic range of the transmit front-end chain \cite{bharadia2013sigcomm,Day2012ITSP}. Denote the covariance of $\boldsymbol{s}_i$ by $\boldsymbol{Q}_i\triangleq\text{Cov}\{\boldsymbol{s}_i\}$. Note that the diagonal elements of $\boldsymbol{Q}_i$ represent the transmit signal power. Following the results in \cite{Santella1998ITVT,Day2012ITSP,Suzuki2008JSAC}, we model $\boldsymbol{e}_i$ as the independent Gaussian vector with zero mean and covariance $\text{Cov}\{\boldsymbol{e}_i\}=\beta \text{diag}(\boldsymbol{Q}_i)$ where $\beta$ is a constant depending on the distortion level of the transmit front-end chain \cite{sahai2013ITVT}.
  
At receiver $i$, the signal of interest $\boldsymbol{h}_{ji}^{\dag}\boldsymbol{s}_{j}, j\neq i$ is received along with the self-interference signal $\boldsymbol{h}_{ii}^{\dag}\boldsymbol{s}_{i}$ and the transmit front-end noise $\boldsymbol{h}_{ji}^{\dag}\boldsymbol{e}_{j}$, $\boldsymbol{h}_{ii}^{\dag}\boldsymbol{e}_{i}$. The power level of $\boldsymbol{h}_{ji}^{\dag}\boldsymbol{e}_{j}, j\neq i$ is typically much lower than that of the thermal noise $N_i$ and thus can be neglected\cite{vehkapera2013PIMRC}. However, $\boldsymbol{h}_{ii}^{\dag}\boldsymbol{e}_{i}$ is in the power level close to the signal of interest and needs to be considered for the analysis, since the gain of the self-interference channel $\boldsymbol{h}_{ii}$ may be $100$dB higher than the gain of the cross-node channel $\boldsymbol{h}_{ji}$ \cite{Day2012ITSP}. In addition to the strength, transmitters and receivers on a same node are relatively static, resulting in long coherence time of self-interference channels, thus receiver $i$ is assumed to have the perfect knowledge of its own self-interference channel $\boldsymbol{h}_{ii}$ \cite{vehkapera2013PIMRC}. Note that receiver $i$ also knows its own transmitted signal $\boldsymbol{s}_i$. Then we can eliminate the self-interference $\boldsymbol{h}_{ii}^{\dag}\boldsymbol{s}_i$ before decoding. The signal after cancellation is given by
\begin{equation}
    Y_i=\boldsymbol{h}_{ji}^{\dag}\boldsymbol{s}_j+\boldsymbol{h}_{ii}^{\dag}\boldsymbol{e}_i+N_i
    \label{Cov of self-interference}
\end{equation} 
where $\boldsymbol{h}_{ii}^{\dag}\boldsymbol{e}_i$ represents the residual self-interference. By defining the aggregate noise term $V_{i}\triangleq\boldsymbol{h}_{ii}^{\dag}\boldsymbol{e}_{i}+N_{i}$, the received signal model can be further simplified as
 \begin{equation}
 \begin{split}
 Y_{i}=\boldsymbol{h}_{ji}^{\dag}\boldsymbol{s}_{j}+V_{i}\\
 \end{split}
  \label{received signal2}
 \end{equation} 
where $V_i$ is Gaussian noise with zero mean and variance $\text{Var}\{V_{i}\}=\sigma_N^2+\beta\boldsymbol{h}_{ii}^{\dag}\text{diag}(\boldsymbol{Q}_{i})\boldsymbol{h}_{ii}$.

\section{Boundary of Achievable Rate Region}
We present the achievable rate region for the MISO full-duplex two-way channel and characterize the boundary points of this region by a family of coupled nonconvex optimization problems. Next, we show that the boundary points can be alternatively obtained by solving a family of convex optimization problems that are the results of the transformation of the original nonconvex optimization problems. 
 
\subsection{Achievable Rate Region and Pareto Boundary}
 As shown in (\ref{received signal2}), the channel model for the wireless link from node $2$ to node $1$ is equivalent to a Gaussian channel. It follows from the results of \cite{telatar1999capacity,cover2012elements} that by employing a Gaussian codebook at node $2$, we can achieve the maximum rate for the channel from node $2$ to node $1$
\begin{equation}
R_1(\boldsymbol{Q}_1,\boldsymbol{Q}_2) =  
 \log \Bigg(
1+\frac{\boldsymbol{h}_{21}^{\dag}\boldsymbol{Q}_2\boldsymbol{h}_{21}}{\sigma_N^2+\beta \boldsymbol{h}_{11}^{\dag}\text{diag}
(\boldsymbol{Q}_1)\boldsymbol{h}_{11}}\Bigg)
\label{maximum rate}
\end{equation}
where $(\boldsymbol{Q}_{1},\boldsymbol{Q}_2)$ are the given transmit covariance matrices. Similarly, the maximum rate for the channel from node $1$ to node $2$ is equal to
\begin{equation}
R_2(\boldsymbol{Q}_1,\boldsymbol{Q}_2) =  
 \log \Bigg(
1+\frac{\boldsymbol{h}_{12}^{\dag}\boldsymbol{Q}_1\boldsymbol{h}_{12}}{\sigma_N^2+\beta \boldsymbol{h}_{22}^{\dag}\text{diag}
(\boldsymbol{Q}_2)\boldsymbol{h}_{22}}\Bigg).
\end{equation}
Any rate pair $(r_1,r_2)$ with $r_1\leq R_1, r_2 \leq R_2$, is achievable for the MISO full-duplex two-way channel. 

Define the achievable rate region for the MISO full-duplex two-way channel to be the set of all achievable rate pairs under the transmit power constraint $P_i$: 
\begin{equation}
\mathcal{R}\triangleq\bigcup_{\scriptstyle\textbf{tr}(\boldsymbol{Q}_i)\leq P_i,
\atop
\scriptstyle \boldsymbol{Q}_i\succeq 0,i=1,2}
\left\lbrace 
  \begin{split}
  &(r_1,r_2):\\
  & 0\leq r_1 \leq R_1(\boldsymbol{Q}_1,\boldsymbol{Q}_2)\\
  & 0\leq r_2 \leq R_2(\boldsymbol{Q}_1,\boldsymbol{Q}_2)
  \end{split}
  \right\rbrace .
\label{centralized region} 
\end{equation}
A rate pair is on the boundary of the rate region $\mathcal{R}$ if and only if it is Pareto optimal which is defined as follows (A similar definition can be found in \cite{Mochaourab2011ITSP,jorswieck2008ITSP,zhang2010ITSP}).

\textit{Definition} 1: (Pareto optimality) A rate pair $(R_1^*,R_2^*)\in \mathcal{R} $ is Pareto optimal
if there does not exist another rate pair $(R_1,R_2)\in \mathcal{R}$ such that $(R_1,R_2)\geq (R_1^*,R_2^*)$ and $(R_1,R_2)\neq (R_1^*,R_2^*)$ where the inequality is component-wise.

Accordingly, the set of all boundary points of the achievable rate region $\mathcal{R}$ is called Pareto boundary and defined as
\begin{equation}
\mathcal{R}^*=\bigcup \;\{\text{all the Pareto optimal rate pairs}\;\; (R_1^*, R_2^*)\}.
\label{Pareto boundary}
\end{equation}
It is shown in \cite{Shang2007Asilomar,Shang2011ITInf} that $\mathcal{R}^*$ can be derived by solving a family of nonconvex optimization problems:
\begin{equation}
\begin{split}
\max\;\;& {\mu_1R_1(\boldsymbol{Q}_1,\boldsymbol{Q}_2)+\mu_2R_2(\boldsymbol{Q}_1,\boldsymbol{Q}_2)}\\
\text{subject to}\;\;
& \textbf{tr}(\boldsymbol{Q}_i)\leq P_i, \boldsymbol{Q}_i\succeq 0, i=1,2
\end{split}
\label{non-convex boundary problems}
\end{equation}
where $0\leq\mu_1<\infty$ and $0\leq\mu_2<\infty$. However, the non-convexity of problem (\ref{non-convex boundary problems}) implies that we must go through all possible transmit covariance matrices $\boldsymbol{Q}_1$ and $\boldsymbol{Q}_2$ to find the optimal solution for each $(\mu_1,\mu_2)$ pair. What is worse, the complexity of such exhaustive search exponentially increases with the dimensions of $\boldsymbol{Q}_1$ and $\boldsymbol{Q}_2$, which renders the computation intractable \cite{Shang2011ITInf}. Next, we present an alternative approach more suitable for deriving the Pareto boundary for the MISO full-duplex two-way channel.


\subsection{Decoupled Optimization Problems}
The difficulty in deriving Pareto boundary $\mathcal{R}^*$ is caused by the non-convexity and the coupled high-dimensional nature of problem (\ref{non-convex boundary problems}). 
To reduce the computationally complexity, we need to decouple problem (\ref{non-convex boundary problems}) in terms of lower-dimensional variables. Using the decoupling procedure in \cite{Shang2007Asilomar,Shang2011ITInf,zhang2010ITSP} we introduce an auxiliary variable $z_i$ for node $i$ which denotes the power of the signal of interest at node $j$ i.e., $z_i\triangleq{\boldsymbol{h}_{ij}^{\dag}\boldsymbol{Q}_i\boldsymbol{h}_{ij}}$. Consider the following optimization problem for node $i$ with the transmit power constraint $P_i$: 
\begin{equation}
\begin{split}
\min\;\;& 
{\boldsymbol{h}_{ii}^{\dag}\text{diag}(\boldsymbol{Q}_i)\boldsymbol{h}_{ii}}\\
\text{subject to}\;\;& {\boldsymbol{h}_{ij}^{\dag}\boldsymbol{Q}_i\boldsymbol{h}_{ij}}=z_i\\
& \textbf{tr}(\boldsymbol{Q}_i)\leq P_i, \boldsymbol{Q}_i\succeq 0
\end{split}
\label{region split problem1}
\end{equation}
where $i,j\in\{1,2\}$ and $i\neq j$. We require 
\begin{equation}
 0\leq z_i\leq\max\limits_{\textbf{tr}(\boldsymbol{Q}_i)\leq P_i, \boldsymbol{Q}_i\succeq 0}\boldsymbol{h}_{ij}^{\dag}\boldsymbol{Q}_i\boldsymbol{h}_{ij}=P_i\|\boldsymbol{h}_{ij}\|^2
 \label{range of z}
\end{equation}
so that problem (\ref{region split problem1}) always has a feasible solution. Denote the optimal value of problem (\ref{region split problem1}) as $\Gamma_i^*(z_i)$. Then, we define a set in terms of $z_i$ and $\Gamma_i^*(z_i)$ as follows:
\begin{equation}
\overline{\mathcal{R}}\triangleq\bigcup_{
\scriptstyle z_1\in[0,P_1\|\boldsymbol{h}_{12}\|^2],
\atop
\scriptstyle z_2\in[0,P_2\|\boldsymbol{h}_{21}\|^2]}
\left\lbrace
\begin{split}
&
(r_1,r_2):\\
& r_1= \log \Bigg(
1+\frac{z_2}{\sigma_N^2+\beta\Gamma_1^*(z_1)}\Bigg)\\
& r_2=\log \Bigg(
1+\frac{z_1}{\sigma_N^2+\beta\Gamma^*_2(z_2)}\Bigg)
\end{split}
\right\rbrace. 
\nonumber   
 \label{convex boundary} 
\end{equation}
Any rate pair in the above set $\overline{\mathcal{R}}$ can be achieved by the optimal solution $\boldsymbol{Q}_1^*$ and $\boldsymbol{Q}_2^*$ for problem (\ref{region split problem1}) with some $z_1$ and $z_2$, thus  $\overline{\mathcal{R}}$ is a subset of the achievable rate region $\mathcal{R}$ in (\ref{centralized region}). In Lemma 1 we show that $\overline{\mathcal{R}}$ includes the Pareto boundary $\mathcal{R}^*$ for the region $\mathcal{R}$.

\textit{Lemma} 1: 
Under the transmit power constraint $P_i$, any point on the Pareto boundary $\mathcal{R}^*$ for the achievable rate region $\mathcal{R}$ in (\ref{centralized region}) can be achieved by the optimal solution $\boldsymbol{Q}_i^*$ for problem (\ref{region split problem1}) with some $z_i$. That is to say, $\mathcal{R}^*\subseteq\overline{\mathcal{R}}$. 
\begin{IEEEproof}
For any point $(R_1^*,R_2^*)$ on the Pareto boundary, assume that it is achieved by $\boldsymbol{Q}_1^*$ and $\boldsymbol{Q}_2^*$. $\boldsymbol{Q}_i^*$ is a feasible solution for problem (\ref{region split problem1}) with $z_i=z_i^*=\boldsymbol{h}_{ij}^{\dag}\boldsymbol{Q}_i^*\boldsymbol{h}_{ij}$ where $i,j\in\{1,2\}$ and $i\neq j$. Let $i=1$, if $\boldsymbol{Q}_1^*$ is not an optimal solution for problem (\ref{region split problem1}) i.e., $\boldsymbol{h}_{11}^{\dag}\text{diag}(\boldsymbol{Q}_1^*)\boldsymbol{h}_{11}>\Gamma_1^*(z_1^*)$ then 
\begin{equation}
R_1^*<\log \Bigg(
1+\frac{z_2^*}{\sigma_N^2+\beta\Gamma_1^*(z_1^*)}\Bigg)=\overline{R_1},
\nonumber
\end{equation}
while
\begin{equation}
R_2^*\leq\log \Bigg(
1+\frac{z_1^*}{\sigma_N^2+\beta\Gamma_2^*(z_2^*)}\Bigg)=\overline{R_2}.
\nonumber
\end{equation}
As $(\overline{R_1},\overline{R_2})$ belongs to $\overline{\mathcal{R}}$ and thus belongs to $\mathcal{R}$, $R_1^*<\overline{R_1}$ and $R_2^*\leq\overline{R_2}$ contradict to the Pareto optimality of $(R_1^*,R_2^*)$. Therefore $\boldsymbol{Q}_1^*$ is an optimal solution for problem (\ref{region split problem1}). In the same way we can show that $\boldsymbol{Q}_2^*$ is an optimal solution for problem (\ref{region split problem1}).
\end{IEEEproof}

We stress that the set $\overline{\mathcal{R}}$ is not necessarily equivalent to the Pareto boundary $\mathcal{R}^*$, since $\overline{\mathcal{R}}$ may include the rate pairs inside the region $\mathcal{R}$. However,  the relationship $\mathcal{R}^*\subseteq\overline{\mathcal{R}}$ implies that any approach of obtaining the set $\overline{\mathcal{R}}$ will suffice to derive the entire Pareto boundary $\mathcal{R}^*$. Furthermore, any result applying to $\overline{\mathcal{R}}$ also works for $\mathcal{R}^*$. Hence, we proceed to explore the optimal signaling for the MISO full-duplex two-way channel by the study of the set $\overline{\mathcal{R}}$.  

\section{Optimal Signaling}
By solving problem (\ref{region split problem1}), we show the optimality of transmit beamforming for the MISO full-duplex two-way channel. In other words, all the points on the Pareto boundary $\mathcal{R}^*$ of the achievable rate region $\mathcal{R}$ can be achieved by transmit beamforming. To better understand the optimal signaling, we provide the closed form of the optimal beamforming weights.

\subsection{Semidefinite Programming Reformulation}
Problem (\ref{region split problem1}) is not a common optimization problem since the objective function includes the non-linear operator $\text{diag}(\cdot)$. By setting $\boldsymbol{A}_i=\boldsymbol{h}_{ij}\boldsymbol{h}_{ij}^\dag$, $\boldsymbol{C}_i=\text{Diag}(|\boldsymbol{h}_{ii}^{(1)}|^{2},\dots,|\boldsymbol{h}_{ii}^{(M)}|^{2})$ and using the equivalent relationship
$\boldsymbol{h}_{ii}^{\dag}\text{diag}(\boldsymbol{Q}_i)\boldsymbol{h}_{ii}=\textbf{tr}(\boldsymbol{C}_i\boldsymbol{Q}_i)$, we reformulate problem (\ref{region split problem1}) to the semi-definite programming (SDP) problem as follows (See more details about SDP in \cite{boyd2009convex}):
\begin{equation}
\begin{split}
\min\;\;& {\textbf{tr}(\boldsymbol{C}_i\boldsymbol{Q}_i)}\\
\text{subject to}\;\;&\textbf{tr}(\boldsymbol{A}_i\boldsymbol{Q}_i)=z_i,\\
&\textbf{tr}(\boldsymbol{Q}_i)\leq P_i,\boldsymbol{Q}_i \succeq 0\\
\end{split}
\label{two node SDP}
\end{equation}
where $\boldsymbol{C}_i, \boldsymbol{A}_i\in \mathbb{H}^M$.

The above SDP reformulation reveals the hidden convexity of problem (\ref{region split problem1}) so that we can solve it by employing the well-developed interior-point algorithm within polynomial time. Furthermore, we can numerically characterize the Pareto boundary for the MISO full-duplex two-way channel in efficiency.

\subsection{Optimal Beamforming}
The optimal solutions for problem (\ref{two node SDP}) determine the signaling structure to achieve the rate pairs in the set $\overline{\mathcal{R}}$. In Theorem 1, we explore the rank of optimal solutions $\boldsymbol{Q}^*_i$ for problem (\ref{two node SDP}) where $i,j\in\{1,2\}$ and $i\neq j$.

\textit{Theorem} 1: For problem (\ref{two node SDP}) with $P_i \geq 0$ and  $0\leq z_i\leq P_i\|\boldsymbol{h}_{ij}\|^2$, there always exists an optimal solution $\boldsymbol{Q}_i^*$  with $\text{rank}(\boldsymbol{Q}^*_i)=1$.

\begin{IEEEproof}
See the Appendix.
\end{IEEEproof}

Note that the transmit signal with the rank-one covariance matrix can be implemented by transmitter beamforming. It follows from Theorem 1 that all points in the set $\overline{\mathcal{R}}$, which include the entire Pareto boundary, can be achieved by the transmitter beamforming. Therefore, we conclude that transmitter beamforming is an optimal scheme for the MISO full-duplex two-way channel. In Lemma 2 we derive the closed-form optimal weights for transmitter beamforming.

\textit{Lemma} 2: For node $i$ in the MISO point-to point full-duplex wireless network with the transmit power constraint $P_i$ and complex channels $\boldsymbol{h}_{ii}, \boldsymbol{h}_{ij}, i,j\in\{1,2\}, i\neq j$, the optimal beamforming weights have the following form:

\begin{equation}
\boldsymbol{w}_i^*=\frac{\sqrt{z_i}(\boldsymbol{C}_i+\epsilon\boldsymbol{I})^{-1}\boldsymbol{h}_{ij}}{\boldsymbol{h}_{ij}^\dag(\boldsymbol{C}_i+\epsilon\boldsymbol{I})^{-1}\boldsymbol{h}_{ij}}
\label{optimal bf}
\end{equation}
where $\boldsymbol{C}_i=\text{Diag}(|\boldsymbol{h}_{ii}^{(1)}|^{2},\dots,|\boldsymbol{h}_{ii}^{(M)}|^{2})$, constant $z_i$ is within the range $0\leq z_i\leq P_i\|\boldsymbol{h}_{ij}\|^2$ and $\boldsymbol{I}$ denotes the $M\times M$ identical matrix. For a fixed $z_i$, nonnegative constant $\epsilon$ is adjusted to satisfy the transmit power constraint $\|\boldsymbol{w}_i\|^2\leq P_i$. Specially, $\epsilon=0$ if
\begin{equation}
 z_i\leq \frac{P_i(\boldsymbol{h}_{ij}^{\dag}\boldsymbol{C}_i^{-1}\boldsymbol{h}_{ij})^2}{\boldsymbol{h}_{ij}^{\dag}\boldsymbol{C}_i^{-2}\boldsymbol{h}_{ij}}.
 \label{low zi condition}
\end{equation} 
\begin{IEEEproof}
The optimal beamforming weights can be obtained by solving problem (\ref{two node SDP}) with the rank-one constraint $\boldsymbol{Q}_i=\boldsymbol{w}_{i}\boldsymbol{w}_{i}^{\dag}$ as follows:
\begin{equation}
\begin{split}
\min\;\;&{\boldsymbol{w}_{i}^{\dag}\boldsymbol{C}_i\boldsymbol{w}_{i}} \\
\text{subject to}\;\;&|\boldsymbol{w}_{i}^{\dag}\boldsymbol{h}_{ij}|^2=z_i, \|\boldsymbol{w}_{i}\|^2\leq P_i.
\end{split}
\label{minimizaing BF problem1}
\end{equation}
 The above problem has the general closed-form optimal solution (\ref{optimal bf}) (see details in \cite{cox1987ITASP}). Without the transmit power constraint $\|\boldsymbol{w}_{i}\|^2\leq P_i$, problem (\ref{minimizaing BF problem1}) has the following optimal solution (shown in \cite{cox1987ITASP})
 \begin{equation}
 \boldsymbol{w}_i^*=\frac{\sqrt{z_i}\boldsymbol{C}_i^{-1}\boldsymbol{h}_{ij}}{\boldsymbol{h}_{ij}^\dag\boldsymbol{C}_i^{-1}\boldsymbol{h}_{ij}}.
 \nonumber
 \end{equation}
 Combining with the condition (\ref{low zi condition}), we obtain 
\begin{equation}
\|\boldsymbol{w}_{i}^*\|^2=\frac{z_i\boldsymbol{h}_{ij}^{\dag}\boldsymbol{C}_i^{-2}\boldsymbol{h}_{ij}}{(\boldsymbol{h}_{ij}^{\dag}\boldsymbol{C}_i^{-1}\boldsymbol{h}_{ij})^2}\leq P_i.
\nonumber
\end{equation} 
Hence, we conclude that $\epsilon=0$ under the condition (\ref{low zi condition}).
\end{IEEEproof}
We remark that the optimal beamforming weights for node $i$ are closely parallel to the cross-node channel $\boldsymbol{h}_{ij}$, beamforming the signal of interest at node $j$. While the transmit front-end noise corresponding to the stronger self-interference channel is largely suppressed via the matrix $(\boldsymbol{C}_i+\epsilon\boldsymbol{I})^{-1}$.

\section{Numerical Examples}

\begin{figure}[!t] 
\centering
\includegraphics[width=3in]{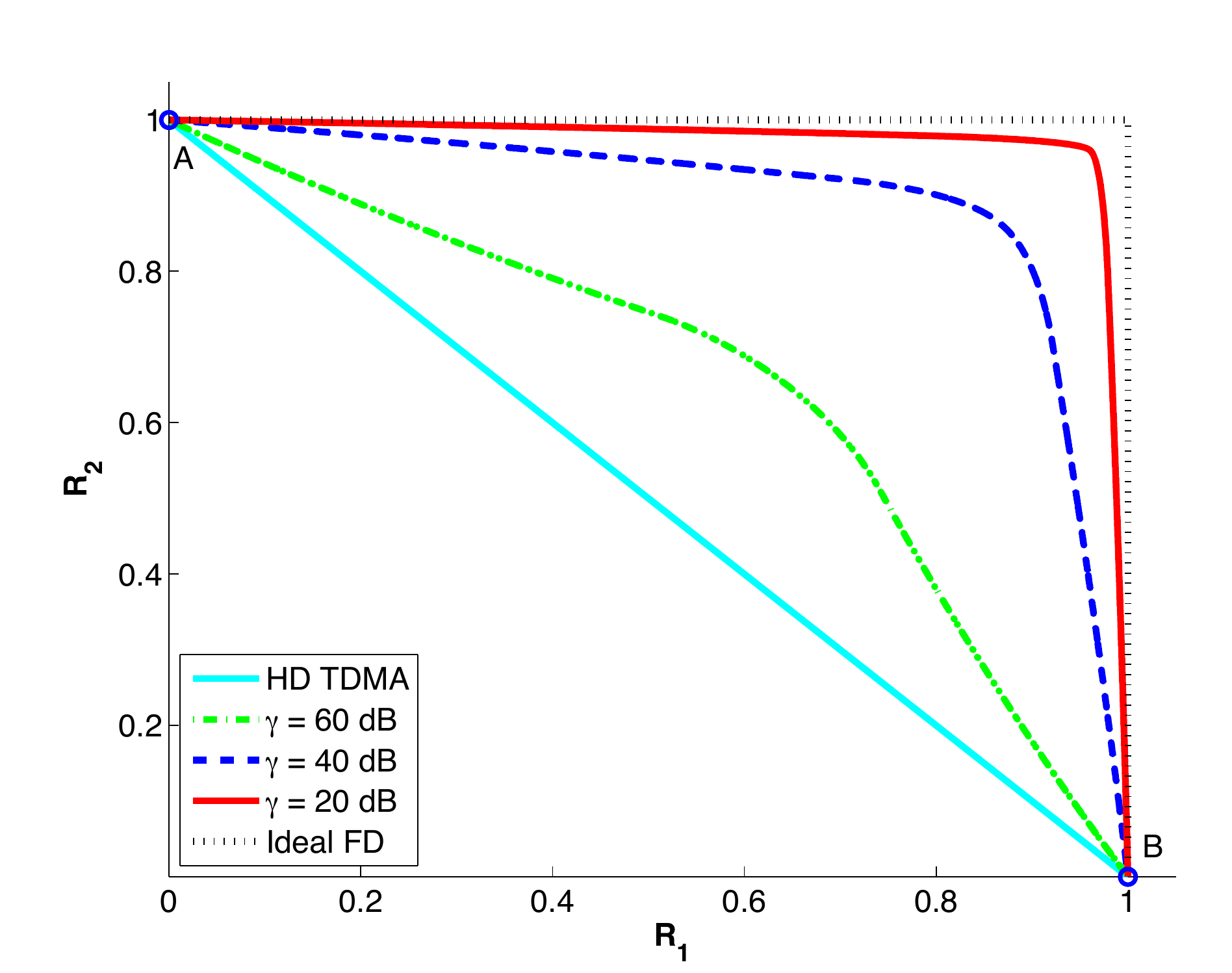}
\caption{The achievable rate regions for the symmetric MISO full-duplex two-way channels with $\beta=-40\;\text{dB}$, $\|\boldsymbol{h}_{12}\|=\|\boldsymbol{h}_{21}\|=1$, $P_1=P_2=1$. As plotted for comparison is the half-duplex TDMA achievable rate region.} 
\label{Fig4}
\end{figure}

We present the achievable rate regions for the MISO full-duplex two-way channels in Fig. \ref{Fig4} where the channels are symmetric i.e., $\boldsymbol{h}_{12}=\boldsymbol{h}_{21}$ and $\boldsymbol{h}_{11}=\boldsymbol{h}_{22}$. Each node is equipped with $M=3$ transmit antennas and single receive antenna with the transmit power constraints $P_1=P_2$ and the receiver noise $\sigma_N^2=1$. Define $\gamma=\frac{\|\boldsymbol{h}_{11}\|}{\|\boldsymbol{h}_{21}\|}=\frac{\|\boldsymbol{h}_{22}\|}{\|\boldsymbol{h}_{12}\|}$ (in dB) to represent the ratio of the self-interference channel gain to the cross-node channel gain. Note that $\gamma$ can be reduced by the passive suppression \cite{Everett2011Asilomar}. The transmit front-end noise level is fixed with $\beta=-40$dB. Each colored line represents the Pareto boundary of the achievable rate region for the channel with corresponding $\gamma$. We conclude from the numerical results that the achievable rate region shrinks as $\gamma$ varies from $20$dB to $60$dB. However, the full-duplex channel always outperforms than the half-duplex TDMA channel if the optimal beamforming is employed. The extreme points $A,B$ of the rate regions on the axes represent the maximum rates in the case that only one-way of the two-way channel is working. It follows that the points $A,B$ are only determined by the transmit power constraints $P_i$. The ideal MISO full-duplex two-way channel sets the outer bound for the achievable rate regions of all channels, doubling that of the half-duplex TDMA channel.

\begin{figure}[!t] 
\centering
\includegraphics[width=2.75in]{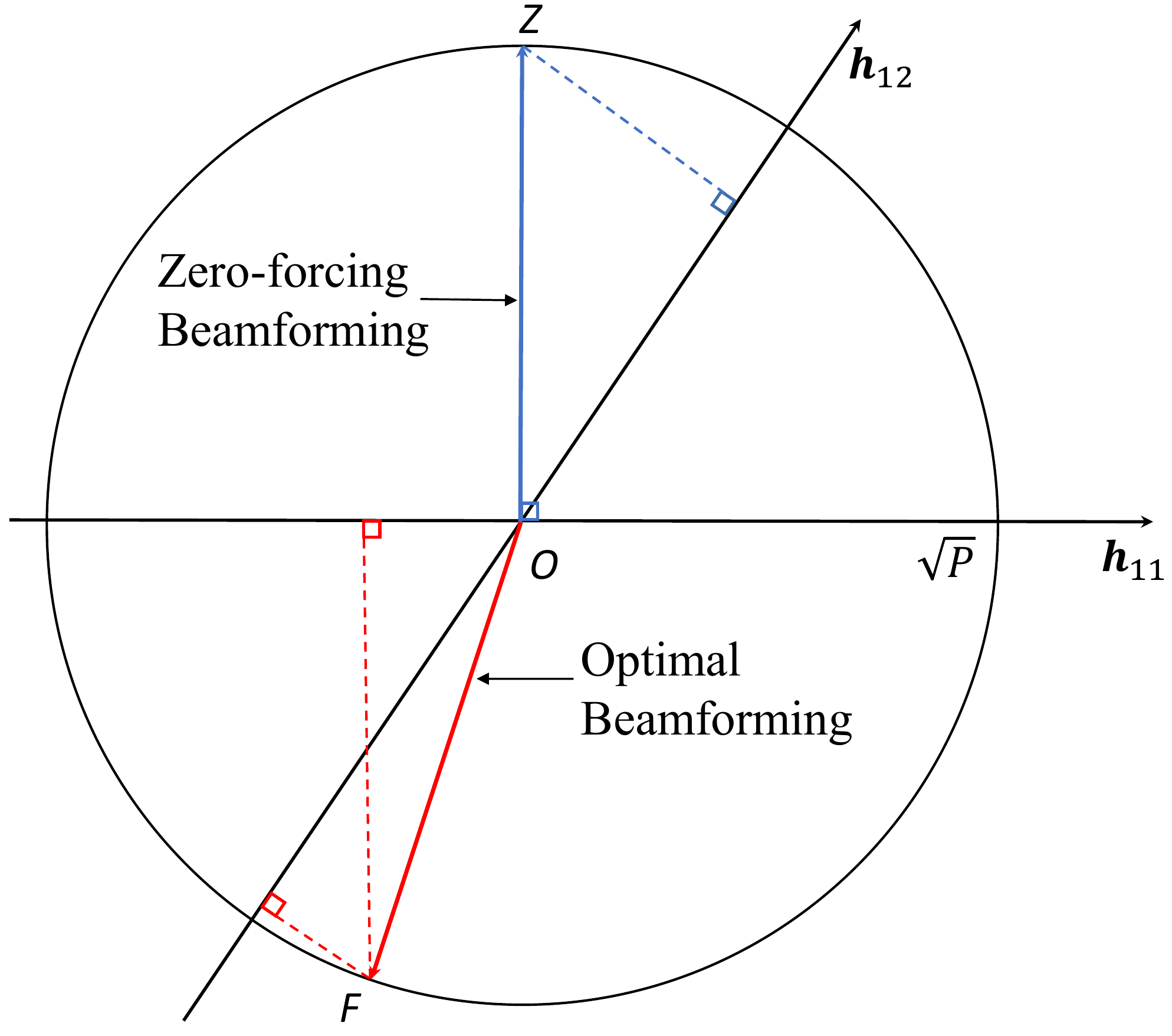}
\vspace{-0.05cm}
\caption{The geometric comparison of the full-duplex optimal beamforming with the zero-forcing beamforming} 
\label{Fig8}
\end{figure}

The rate pair with the maximum sum-rate corresponds to the point with $R_1=R_2$ on the Pareto boundary and can be achieved by certain weights in the set of optimal beamforming weights defined in Lemma 2. In Fig. \ref{Fig8} we compare the optimal weights  $\overline{OF}$ which corresponds to the maximum sum-rate with the zero-forcing (ZF) beamforming weights $\overline{OZ}$. For simplicity, we assume all channel vectors to be $2\times 1$ real vectors with  $\boldsymbol{h}_{11}=\boldsymbol{h}_{22}$ and $\boldsymbol{h}_{12}=\boldsymbol{h}_{21}$. Assume the transmit power constraints $P_1=P_2=P$. Then all possible beamforming weights are contained in the disc with radius $\sqrt{P}$. $\overline{OZ}$ restricts the transmit signal orthogonal to the self-interference channel $\boldsymbol{h}_{ii}$, whereas $\overline{OF}$ is not orthogonal to $\boldsymbol{h}_{ii}$ but has greater length of projection on the cross-node channel $\boldsymbol{h}_{ij}$. 

\begin{figure}[!t] 
\centering
\includegraphics[width=3in]{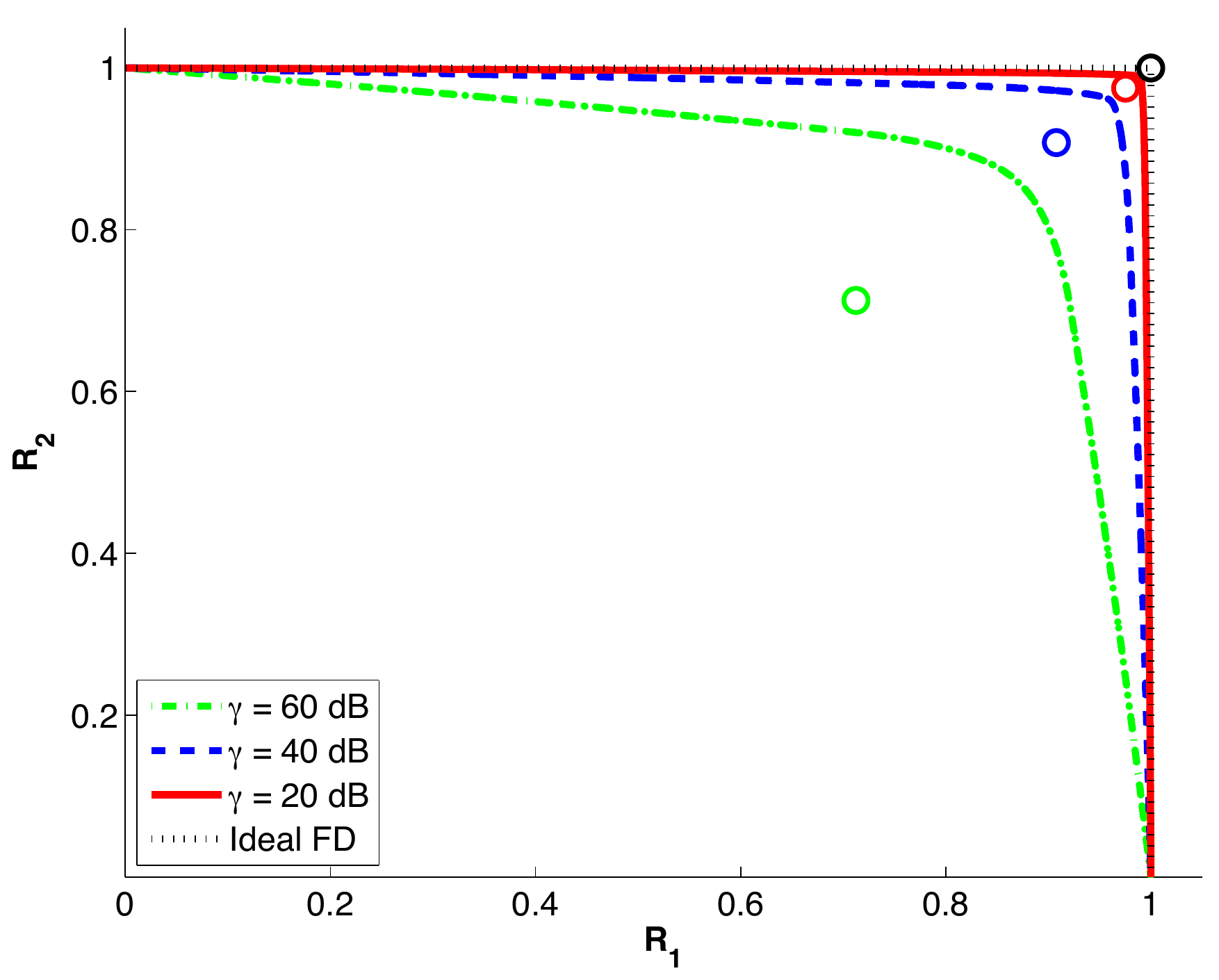}
\caption{The achievable rate regions for the symmetric MISO two-way full-duplex channel with $\beta=-60\;\text{dB}$. Circles denote the respective ZF beamforming rates for $\gamma=20\;\text{dB},\gamma=40\;\text{dB},\gamma=60\;\text{dB}$ and the ideal full-duplex network.} 
\label{Fig6}
\end{figure}

In Fig. \ref{Fig6}, we evaluate the performance of the ZF beamforming for the same full-duplex channels as in Fig. \ref{Fig4} but with $\beta=-60$ dB. Comparing the channel with the same $\gamma$ in Fig. \ref{Fig4} and Fig. \ref{Fig6}, the achievable rate region is increased due to reduction
of $\beta$. The circles, which represent the rate pairs achieved by the ZF beamforming, are below the corresponding Pareto boundaries except for the ideal full-duplex channel. It follows that the ZF beamforming is not optimal for the MISO full-duplex channel in presence of the residual self-interference. As shown in Fig. \ref{Fig8}, the ZF beamforming generates no interference at the unintended receiver if the interference equals to the projection of the transmit signal on the interference channel. However, for the full-duplex case in (\ref{Cov of self-interference}), the residual self-interference signal statistically depends on the transmit signal power rather than being the projection of the transmit signal on $\boldsymbol{h}_{ii}$. Therefore, the ZF beamforming is inefficient in the suppression of the residual self-interference. With $\gamma$ decreasing, the residual self-interference signal is gradually weaker and thus the ZF beamforming is closer to the optimal beamforming and is exactly optimal in the ideal full-duplex network.

\section{Conclusion}
We considered the MISO point-to-point full-duplex wireless network. We derived the achievable rate region and the characterization of the Pareto boundary for the MISO two-way full-duplex channel in presence of the transmit front-end noise. Using the decoupling technique and SDP reformulation, we proposed a new method to obtain the entire Pareto boundary by solving a family of convex SDP problems, rather than the original non-convex problems. We showed that any rate pair on the Pareto boundary can be achieved by the beamforming transmission strategy. Finally, we provided the closed-form solution for the optimal beamforming weights of the MISO full-duplex two-way channel. 

\section*{Appendix}
\textit{Proof of Theorem 1}:
We prove Theorem 1 by the primal-dual method. Note that problem (\ref{two node SDP}) is feasible and bounded. It follows that its dual problem is also feasible and bounded \cite{boyd2009convex}. Assume $\boldsymbol{Q}^*$ is an optimal solution for problem (\ref{two node SDP}). From \cite{boyd2009convex}, problem (\ref{two node SDP}) has the dual problem as follows:
\begin{equation}
\begin{split}
\min_{\lambda_1,\lambda_2}\;\;&{\lambda_1z_i+\lambda_2P} \\
\hspace{20pt}\text{subject to}\;\;&\boldsymbol{Z}=\boldsymbol{C}_i-{\lambda_1\boldsymbol{A}_i}-{\lambda_2\boldsymbol{I}}\succeq 0.\\
\end{split}
\label{equality SDP Dual}
\end{equation} 
where $P=\textbf{tr}(\boldsymbol{Q}^*)$. Assume $((\lambda_1^*,\lambda_2^*),\boldsymbol{Z}^*)$ are the optimal solutions for (\ref{equality SDP Dual}). We denote the rank of 
$\boldsymbol{Q}^*$ by $r$. We assume $r>1$. Following that $\boldsymbol{Q}^*$ is positive semi-definite, $\boldsymbol{Q}^*$ can then be written as $\boldsymbol{Q}^*=\boldsymbol{V}\boldsymbol{V}^{\dag}$ via the singular-value decomposition where $\boldsymbol{V}\in \mathbb{C}^{M \times r}$.
 
Next, we consider the following two linear equations defined by $\boldsymbol{A}_i$ and $\boldsymbol{V}$:
\begin{equation}
\left\lbrace 
\begin{split}
&\textbf{tr}(\boldsymbol{V}^{\dag}\boldsymbol{A}_i\boldsymbol{V}\boldsymbol{X})=0\\
&\textbf{tr}(\boldsymbol{V}^{\dag}\boldsymbol{V}\boldsymbol{X})=0\\
\end{split}
\right. 
\label{linear system}
\end{equation}  
where the unknown matrix $\boldsymbol{X} \in \mathbb{H}^r$ contains $r^2$ real-valued unknowns, that is, $\frac{r(r+1)}{2}$ for the real part and $\frac{r(r-1)}{2}$ for the imaginary part. 

The linear system (\ref{linear system}) must have a non-zero solution, denoted by $\boldsymbol{X}^*$, since it has $r^2$ unknowns where $r\geq 2$. By decomposing the Hermitian matrix ${\boldsymbol{X}^*}$, we obtain ${\boldsymbol{X}^*}=\boldsymbol{U}\boldsymbol{\Sigma}\boldsymbol{U}^{\dag}$, where $\boldsymbol{U}$ is an $r$ dimensional unitary matrix and $\boldsymbol{\Sigma}$ is the diagonal matrix,  $\boldsymbol{\Sigma}=\text{Diag}(\sigma_1,\dots,\sigma_r)$. Without loss of generality, we assume $|\sigma_1|\geq|\sigma_2|\dots\geq|\sigma_r|$. Non-zero matrix $\boldsymbol{X}^*$ has at least one non-trivial eigenvalue, thus $|\sigma_1|>0$. Next, we construct a new matrix as follows:
\begin{equation}
{\boldsymbol{Q}^*_{(1)}}=\boldsymbol{V}(\boldsymbol{I}-\frac{1}{\sigma_1}{\boldsymbol{X}^*})\boldsymbol{V}^{\dag}.
\end{equation}
Note that $\boldsymbol{I}-\frac{1}{\sigma_1}{\boldsymbol{X}^*} \succeq 0$. It follows that $\boldsymbol{Q}^*_{(1)}$ is positive semi-definite. Next, we show that $\boldsymbol{Q}^*_{(1)}$ is also an optimal solution for problem (\ref{two node SDP}). Note that $\boldsymbol{Q}^*_{(1)}$ is optimal for problem (\ref{two node SDP}) if and only if $(\boldsymbol{Q}^*_{(1)}, (\lambda_1^*,\lambda_2^*),\boldsymbol{Z}^*)$ satisfies the KKT conditions, including the primal feasibility, the dual feasibility and the complementarity \cite{Huang2010ITSP}.
As $((\lambda_1^*,\lambda_2^*),\boldsymbol{Z}^*)$ is unchanged, the dual feasibility is automatically satisfied. Therefore, we need only to prove the primal feasibility and the complementarity of $\boldsymbol{Q}^*_{(1)}$.

$\boldsymbol{Q}^*_{(1)}$ is a feasible solution for problem (\ref{two node SDP}), since the following two equations hold for $\boldsymbol{Q}^*_{(1)}$, 
\begin{equation}
\begin{split}
\textbf{tr}(\boldsymbol{A}_i\boldsymbol{Q}^*_{(1)})&=\textbf{tr}(\boldsymbol{A}_i\boldsymbol{V}(\boldsymbol{I}-\frac{1}{\sigma_1}{\boldsymbol{X}^*})\boldsymbol{V}^{\dag})\\
&=\textbf{tr}(\boldsymbol{V}^{\dag}\boldsymbol{A}_i\boldsymbol{V}\boldsymbol{I})-\frac{1}{\sigma_1}\textbf{tr}(\boldsymbol{V}^{\dag}\boldsymbol{A}_i\boldsymbol{V}{\boldsymbol{X}^*})=z_i,
\end{split}
\nonumber
\end{equation}
\vspace{-0.3cm}
\begin{equation}
\begin{split}
\textbf{tr}(\boldsymbol{Q}^*_{(1)})&=\textbf{tr}(\boldsymbol{V}(\boldsymbol{I}-\frac{1}{\sigma_1}{\boldsymbol{X}^*})\boldsymbol{V}^{\dag})\\
&=\textbf{tr}(\boldsymbol{V}^{\dag}\boldsymbol{V}\boldsymbol{I})-\frac{1}{\sigma_1}\textbf{tr}(\boldsymbol{V}^{\dag}\boldsymbol{V}{\boldsymbol{X}^*})=P.
\end{split}
\nonumber
\end{equation}

To show the complementarity, note that $\textbf{tr}(\boldsymbol{Q}^*\boldsymbol{Z}^*)=\textbf{tr}(\boldsymbol{V}^{\dag}\boldsymbol{Z}^*\boldsymbol{V})=0$ and $\boldsymbol{V}^{\dag}\boldsymbol{Z}^*\boldsymbol{V}\succeq 0$ implies that $\boldsymbol{V}^{\dag}\boldsymbol{Z}^*\boldsymbol{V}=0$. It follows that
 \begin{multline}
 \textbf{tr}(\boldsymbol{Q}^*_{(1)}\boldsymbol{Z}^*)=\textbf{tr}(\boldsymbol{V}(\boldsymbol{I}-\frac{1}{\sigma_1}{\boldsymbol{X}}^*)\boldsymbol{V}^{\dag}\boldsymbol{Z}^*)\\=\textbf{tr}((\boldsymbol{I}-\frac{1}{\sigma_1}{\boldsymbol{X}^*})\boldsymbol{V}^{\dag}\boldsymbol{Z}^*\boldsymbol{V})=0.
 \nonumber
 \end{multline}
Therefore, $\boldsymbol{Q}^*_{(1)}$ is an optimal solution for problem (\ref{two node SDP}). Furthermore, the rank of $\boldsymbol{Q}^*_{(1)}$ is strictly smaller than $r$ since $\text{rank}(\boldsymbol{Q}^*_{(1)})=\text{rank}(\boldsymbol{I}-\frac{1}{\sigma_1}{\boldsymbol{X}^*})<r$. 

We can repeat this process as $\boldsymbol{Q}^*,\boldsymbol{Q}^*_{(1)},\boldsymbol{Q}^*_{(2)},\cdots$, until $\text{rank}(\boldsymbol{Q}^*_{(k)})\leq \sqrt{2}$. In other words, the rank of the optimal solution can be strictly decreasing to $\text{rank}(\boldsymbol{Q}^*_{(k)})\leq \sqrt{2}$, that is, $\text{rank}(\boldsymbol{Q}^*_{(k)})=1$.


%



\bibliographystyle{IEEEtran}
\bibliography{IEEEabrv,Allerton2014}
\end{document}